\documentclass[aps, prb, reprint, amssymb, amsfonts, longbibliography]{revtex4-1}

\usepackage{amsmath}
\usepackage{bm}
\usepackage{graphicx}
\usepackage{bm}
\usepackage[unicode]{hyperref}
\usepackage[latin1]{inputenc}
\usepackage{SIunits}

\begin{document}

\title{A method for mechanical generation of radio frequency fields in nuclear magnetic resonance force microscopy}

\author{J. J. T. Wagenaar}
	\email{Wagenaar@physics.leidenuniv.nl}
\author{A. M. J. den Haan}
\author{R. J. Donkersloot}
\author{F. Marsman}
\author{M. de Wit}
\author{L. Bossoni}
\author{T. H. Oosterkamp}

\affiliation{Kamerlingh Onnes Laboratory, Leiden University, PO Box 9504, 2300 RA Leiden, The Netherlands}

\date{\today}

\begin{abstract}
We present an innovative method for magnetic resonance force microscopy (MRFM) with ultra-low dissipation, by using the higher modes of the mechanical detector as radio frequency (rf) source. This method allows MRFM  on samples without the need to be close to an rf source. Furthermore, since rf sources require currents that give dissipation, our method enables nuclear magnetic resonance experiments at ultra-low temperatures. Removing the need for an on-chip rf source is an important step towards a MRFM which can be widely used in condensed matter physics.
\end{abstract}

\maketitle

\section{Introduction}
Magnetic Resonance Force Microscopy is a technique that enables nuclear magnetic resonance experiments on the nanoscale\cite{Poggio2010}.  MRFM demonstrated three dimensional imaging with ($5$ nm$)^3$ resolution of a tobacco virus\cite{Degen2009}. The technique is also exploited differently, namely in measuring spin-lattice relaxation times \cite{Isaac2016, Alexson2012}. In both kind of measurements, there is the need of an rf source in order to manipulate the spins. This limits the experiment in two ways: the first is the heating that occurs when large rf pulses are required \cite{Poggio2007}, the second limitation is the need for the rf source  to be close to the sample under study, which increases the complexity of the experimental setup and sample preparation. In this paper, we present a method for mechanical generation of rf fields \footnote{A. M. J. den Haan, J. J. T. Wagenaar, and T. H. Oosterkamp, A Magnetic Resonance Force Detection Apparatus and Associated Methods, United Kingdom Patent No. GB 1603539.6 (1 Mar 2016), patent pending}, using the force sensor itself to generate the rf fields required for measuring the nuclear spin-lattice relaxation time in a copper sample, down to temperatures of $42$ mK. This achievement brings MRFM one step closer to being a technique that can be more widely used in condensed matter physics or for three dimensional imaging in biological systems at temperatures well below $1$ K. 

\begin{figure}[tb]
\includegraphics[width=\columnwidth]{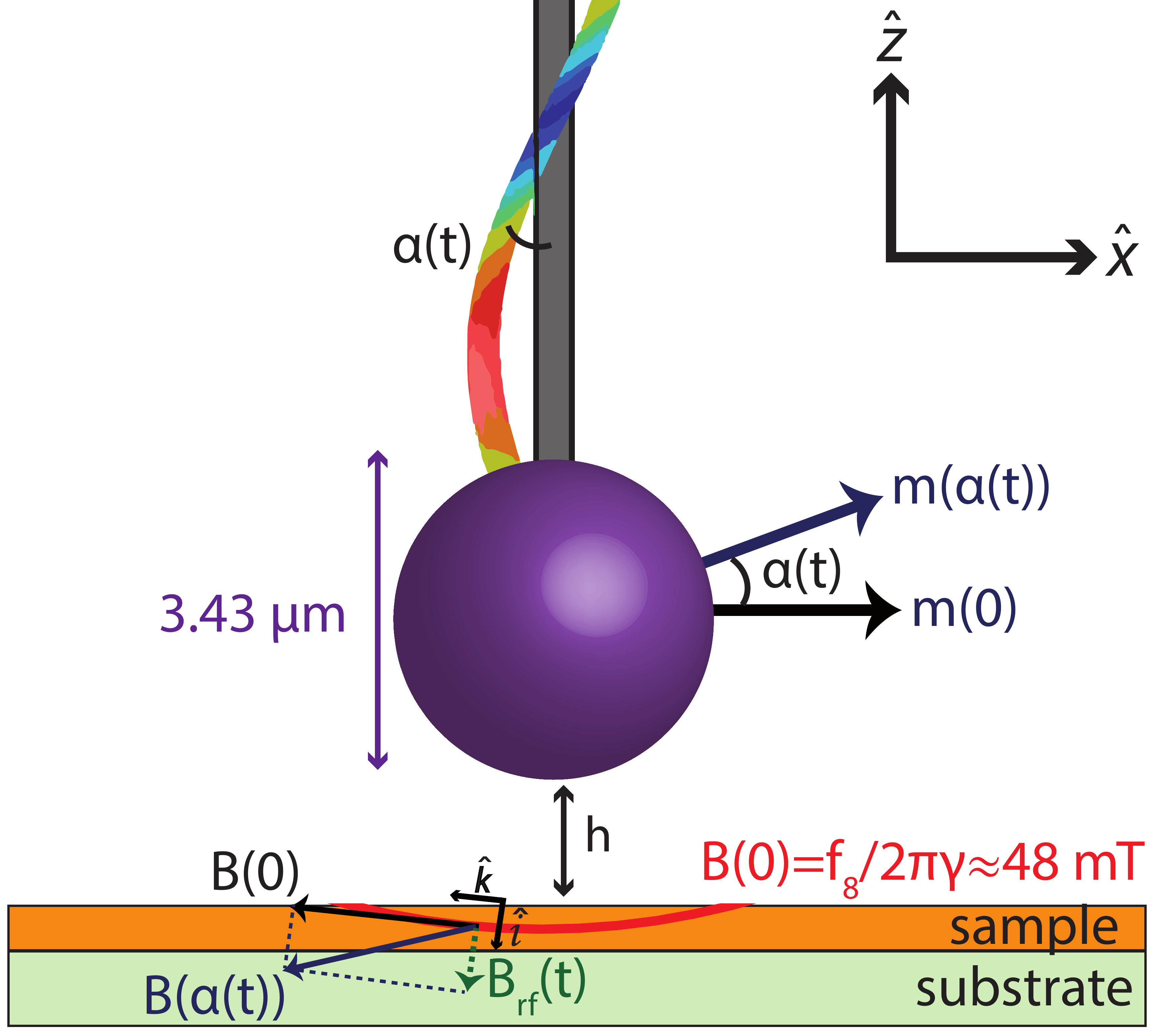}
\caption{Sketch of method to mechanically generate rf fields using the higher resonant modes of the cantilever. A higher mode (rainbow beam) of the cantilever causes a rotation of the magnetic particle with an angle $\alpha(t)$. This rotation effectively results in a magnetic field component $B_{rf}(t)$ perpendicular to the initial magnetic field $B(0)$. Only spins that are positioned within the resonant slice, for which $B(0)=f_n / 2\pi\gamma$, will be perturbed by this rf field. For the $8^{\text{th}}$ mode, $B(0)$ corresponds to $48$ mT and a height $h=1.54$ $\micro$m. }
\label{fig:fig1}
\end{figure}

MRFM makes use of a force sensor, also refereed to as the cantilever, that couples magnetically to electronic or nuclear spins due to a large field gradient. Using magnetic resonance, the spins are manipulated, such that forces or frequency shifts can be measured by the force sensor. The sensitivity of this technique for the purpose of imaging a magnetic moment $\micro_{min}$ is limited by the force noise on the sensor,  and the force generated by the field gradients: ${\micro_{min}=\frac{1}{\nabla B}\sqrt{4k_BT\gamma f_{BW}}}$. Here $\gamma$ is the friction of the cantilever, currently limited by non-contact friction caused by dielectric \cite{Kuehn2006} and magnetic dissipation \cite{Haan2015}. The field gradient $\nabla B$ is generated by a micrometer sized magnetic particle, which is typically $10^5-10^6$ T/m and can be further increased using an even smaller magnetic particle\cite{Overweg2015}. The bandwith of the measurements is given by $f_{BW}$. And $T$ is the temperature of the cantilever, which can be cooled and detected down to $25$ mK using superconducting quantum interference device (SQUID) based read-out \cite{Usenko2011}. The experiments performed on the tobacco virus were performed down to a temperature of the dilution refrigerator of $300$ mK limited by the dissipation in the rf copper micro wire, but with higher temperatures of the cantilever and sample caused by laser heating\cite{Poggio2007, Degen2009}. 

Our experimental setup and sample have been used recently to show that the nuclear spin-lattice relaxation time ($T_1$) of a copper sample can be probed at a nanoscopic scale \cite{Wagenaar2016} at spin temperatures as low as $42$ mK. We used an rf wire to saturate $^{63}$Cu and $^{65}$Cu spins, causing a detectable frequency shift on the magnetic cantilever. In this experiment we found for certain rf frequencies that the number of saturated spins increased, suggesting the presence of an additional rf field. In this paper, we show experimentally that the higher rf fields were generated mechanically by exciting higher modes of the magnetically tipped cantilever. Moreover, we report that for millisecond pulse lengths, the higher modes give rf fields much larger than can currently be obtained using the rf wire solely. When combined with an rf wire, this new technique reduces the dissipation of the conventional rf source, since less rf current is now required to obtain strong enough rf fields for MRFM experiments. When using smaller magnets, we believe that the generated rf fields can be at least $50$ times larger than what is shown here, up to a few mT. 

\section{The idea}
In the case of magnet-on-cantilever geometry where the magnetic particle is attached to the cantilever, the sample experiences large magnetic field gradients and a constant magnetic field $B_0$. Conventionally, it is the magnetic field gradient that results in a shift of the cantilever's (first mode) stiffness $\Delta k_1$\cite{Voogd2015}, which results in a shift in the resonance frequency $\Delta f_1=\frac{1}{2}\frac{f_0\Delta k_1}{k_1}$. Typically, the first mode of the resonator is excited, because the sensitivity in stiffness shift is proportional to the softness of the cantilever. For the first mode, $k_1=7.0\times10^{-5}$ N/m. However, there are other available resonant modes. In Magnetic Force Microscopy for example, torsional modes are used to improve the resolution and reduce the topography-related interference\cite{Kaidatzis2013}. Also it has been proposed that torsional modes can be used to directly couple to a single nuclear spin inside a ferromagnet attached to a cantilever \cite{Butler2010}. We will show how higher modes can be used in MRFM as well, in order to mechanically generate rf fields.

\begin{figure}[tb]
\includegraphics[width=\columnwidth]{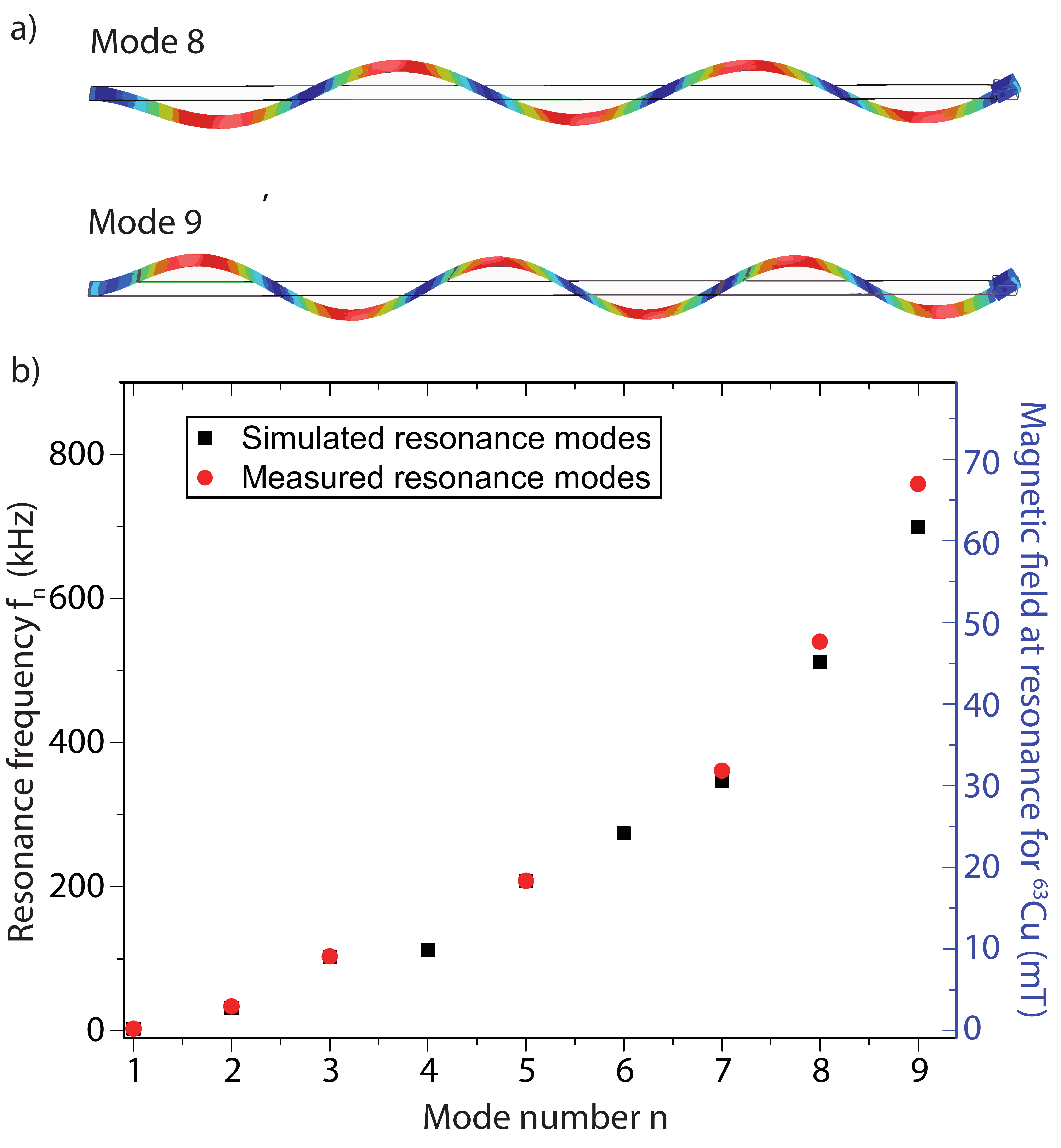}
\caption{\textbf{a)} Simulation of the $8^{\text{th}}$ and $9^{\text{th}}$ mode of the cantilever. These two modes are used in the experimental part of this paper. The dynamics of the magnetic particle at higher modes are mainly rotational, which is due to the relatively larger mass of the magnetic particle compared to the bare silicon cantilever. \textbf{b)} Frequency of the simulated (black) and measured (red) modes of the cantilever. Modes $4$ and $6$ do not oscillate in the x-direction and could therefore not be detected. On the right vertical axis the corresponding magnetic field at resonance ($f_n=\frac{\gamma}{2\pi}B_0$) for the $^{63}$Cu is shown.}
\label{fig:fig2}
\end{figure}

We propose that while the first mode, in our case at $3$ kHz, is used for detection, one can use the higher modes to drive and generate an oscillating field. In figures \ref{fig:fig1} and \ref{fig:fig2}a we sketch the idea of how the motion of higher modes results in a very small rotating magnetic field at the position of the spin, which can be decomposed in a static field and an oscillating rf field.

The idea is as follows: The magnetic particle with magnetic moment $\bm{m}$ gives a magnetic field $B(\bm{r}, t)$ on the spin's location $\bm{r}$. For a spherical particle this is a dipole magnetic field:
\begin{align}
\bm{B}(\bm{r}, t)=\frac{\micro_0}{4\pi}\left(\frac{3\bm{r}(\bm{m}(t)\cdot\bm{r})}{r^5}-\frac{\bm{m}(t)}{r^3}\right)
\end{align} 
When the force sensor is at rest, the magnetic moment of the cantilever is oriented in the $x$-direction, along the direction of motion of the fundamental mode. But when a higher mode with frequency $f_n$ is driven, the particle starts to rotate with an angle $\alpha(t)=\alpha_0 \sin{(2\pi f_{n} t)}$, see figure \ref{fig:fig1}. Assuming that the magnetic particle rotates at a frequency $f_n$, we can write the magnetization and the magnetic field as follows:
\begin{align}
\bm{m}(t)&=m\cos{(\alpha(t))}\bm{\hat{x}}+m\sin{(\alpha(t))}\bm{\hat{z}}\\
&\approx m \bm{\hat{x}}+ \alpha_0 m\sin{(2\pi f_{n} t)}\bm{\hat{z}}
\end{align}
Here we used in the last line that the rotation angle is very small. The magnetic field $B(\bm{r}, \alpha(t))$ will only deviate a bit from the rest position, and thereby the orientation of the spin will be almost constant. However, we know from conventional NMR that a very small perpendicular rf field with the right frequency can cause large perturbations in the spin's alignment when the field oscillation is resonant with the Larmor frequency\cite{Abragam1961}. The perpendicular component of the magnetic field with respect to the magnetic field at $t=0$ can be seen as such a perpendicular perturbing rf field $B_{rf}(\bm{r}, \alpha(t))$:
\begin{align}
|\bm{B}_{rf}(\bm{r}, \alpha(t))|=\frac{|\bm{B}(\bm{r}, \alpha(t))\times \bm{B}(\bm{r}, 0)|}{|\bm{B}(\bm{r}, 0)|}
\label{eq:RF}
\end{align} 
$B_{rf}\propto\alpha$ for small $\alpha$. Neglecting the small parallel component, we can write the magnetic field $\bm{B}(\bm{r}, t)$ as:
\begin{align}
\bm{B}(\bm{r}, t)=|\bm{B}(\bm{r}, 0)|\bm{\hat{k}}+|\bm{B_{rf}}(\bm{r}, \alpha_0)| \sin{(2\pi f_{n} t)}\bm{\hat{i}}
\end{align}
In the local coordinate frame of the spin, with $\bm{\hat{k}}$ parallel to $\bm{B}(\bm{r}, 0)$ and $\bm{\hat{i}}$ chosen parallel to $\bm{B_{rf}}(\bm{r}, \alpha(t))$. This is again conventional NMR with an alternating rf field $B_{rf}\equiv |\bm{B_{rf}}(\bm{r}, \alpha_0)|$, giving resonance when the frequency of the rotation $f_n$ is equal to the Larmor frequency of the spins $f_L=\frac{\gamma}{2\pi}B_0$. In our case, we work with a magnet attached to the cantilever, but we believe the idea holds also for a MRFM with sample-on-tip geometry, when the sample is placed on a node of the cantilever's higher mode motion\cite{Degen2009}.

\section{Experiment}
\subsection{Setup}
The setup consists of the force detector and a copper sample sputtered on top of a detection chip. A detailed description of the setup can be found in references \onlinecite{Wagenaar2016,Haan2015}. Here we will only give a short summary. 
The force sensor is a very soft IBM-type silicon cantilever, with $k_1=7.0\times10^{-5}$ N/m, and length, width and thickness of $145$ $\micro$m, $5$ $\micro$m and $100$ nm respectively\cite{Chui2003,Mamin2003}. Using electron beam induced deposition, a spherical NdFeB magnetic particle with a radius of $1.72$ $\micro$m and a saturation magnetization of $1.15$ T is glued to the end of the cantilever, resulting in a fundamental resonance frequency of $f_1=3.0$ kHz, and a mechanical quality factor of $Q_1=\frac{k_1}{2\pi f_1 \gamma}=2.8\cdot10^4$ at cryogenic temperatures. The cantilever can be moved on a range of $1$ mm using an in-house-developed cryopositioning system, while measuring the absolute position using capacitive sensors. A fine stage below the sample can be used to scan the sample in a range of $2.3$ $\micro$m. The cantilever is driven using a piezo element \footnote{Piezoceramic Tube 5A by EBL Products, Inc., USA; PICMA Chip Monolithic Multylayer Piezo Actuator PL033.30 by Physik Instrumente (PI) GmbH \& Co., Germany.}.

The detection chip consists of a superconducting pick-up loop, $30\times30$ $\micro$m$^2$, connected to a superconducting quantum interference device (SQUID). The motion of the magnetic particle results in a flux change in the pick-up coil, measured by the SQUID\cite{Usenko2011}. Close to the pick-up coil, copper is sputtered on an area of $30\times30$ $\micro$m$^2$ with a thickness of $300$ nm and a roughness of $10$ nm. A gold capping layer is sputtered in order to prevent oxidation of the copper. The copper sample is thermalized to the mixing chamber of the dilution refrigerator using a patterned copper wire leading to a large copper area, which is connected to the sample holder via gold wire bonds. The sample holder is connected via a welded silver wire to the mixing chamber. A superconducting (Nb,Ti)N rf wire, $2$ $\micro$m wide, $300$ nm thick, is positioned from the copper sample at distance of $500$ nm. For the experiments in this paper, we position the cantilever above the copper, close to the rf line ($7\pm1$ $\micro$m) and close to the pickup coil ($5\pm1$ $\micro$m). 

\subsection{Higher modes}
Besides the cantilever's fundamental resonant mode in the $x$-direction, at $f_1=3.0$ kHz, the cantilever exhibits many higher modes. We simulated the eigenmodes using a finite element analysis (COMSOL Multiphysics) with an eigenfrequency study. The results of the mode shapes for the $8^{\text{th}}$ and the $9^{\text{th}}$ mode are visible in figure \ref{fig:fig2}a. Some modes, 4 and 6, are motions that do not oscillate in the x-direction, which give too small flux change in the pickup coil to be detected. We were able to detect the higher modes by driving the cantilever using the rf wire, as well as by using the piezo located on the holder of the cantilever. Figure \ref{fig:fig2}b shows a graph of the simulated and the measured resonance frequencies of the cantilever. 	
		
Most of the higher modes are similar to the $8^{th}$ and $9^{th}$ modes, but with less or more nodes. The node present at the position of the magnetic particle results in a rotation of the magnetic moment $\bm{m}$. In this paper, we will use only modes $8$ and $9$. The even higher modes have resonant slices corresponding to a shorter distance. The eddy currents in the copper prevented us from measurements closer than approximately $1$ $\micro$m distance.  The lower modes have the disadvantage that the corresponding signals would be much less, since the required distance would be larger. We believe that the method described in this paper can also be applied to other modes with a rotation of the magnetic particle, up to the MHz regime. These modes would be available when studying a less dissipative sample.
		
\subsection{Measurement protocol}
In this paper, we initially follow the same measurement protocol as used when the rf wire is used as an rf source. However, by applying shorter pulses with the rf wire, and by applying pulses with a piezo element attached to the cantilever holder, we show that we can excite the higher modes of the cantilever, resulting in a larger magnetic field strength as is obtained by using only the rf wire.

When an rf field $\bm{B_{rf}}$ is applied perpendicular to the direction of the magnetic field which is parallel to the spin's initial orientation, the spin will rotate along the axis of rotation parallel to $\bm{B_{rf}}$ \cite{Abragam1961}. For large rf fields and sufficiently long pulses, the spin transitions saturate, resulting in a netto zero magnetization for ensembles of spins. Every spin whose transitions are saturated, causes a stiffness shift and therefore a frequency shift $-\Delta f_s$, as described above. Only the spins that meet the resonance condition $f_{rf}=\frac{\gamma}{2\pi} B_0$, with $\gamma$ the gyromagnetic ratio, will be excited, thus forming the resonant slice.  The total frequency shift $-\Delta f_1$ can be obtained by summation of all spins within the resonant slice: 
\begin{align}
\Delta f_1=-p\sum_V \Delta f_s(\bm{r})
\label{eq:summation}
\end{align}
Here $p$ is the net Boltzmann polarization of the nuclear spins. 

We drive the cantilever at its resonance frequency $f_1$ using a phase-locked loop (PLL) of a Zurich Instrument lock-in amplifier to measure the frequency shifts. During the pulse, the PLL is switched off, driving the cantilever at it's last known resonance frequency, to avoid possible cross-talk effects. The small frequency shifts and the cantilever's low Q-factor enable the PLL to drive well within the cantilever's line-width during the pulse, thereby keeping the oscillation amplitude of the cantilever constant.

After a saturation pulse, the nuclear spins will exponentially restore their initial orientation in a characteristic time $T_1$. For copper, the relaxation time follows the Korringa relation\cite{Korringa1950} $TT_1=1.2$ sK, which is confirmed in the experiments with the rf wire as the rf source\cite{Wagenaar2016}. After a saturation pulse ending at $t=0$ , the frequency shift evolves as function of time $\Delta f(t)$ according to:
\begin{align}
\Delta f(t)=-\Delta f_1 \cdot e^{-t/T_1}
\label{eq:fitting}
\end{align}			
\begin{figure}[tb]
\includegraphics[width=\columnwidth]{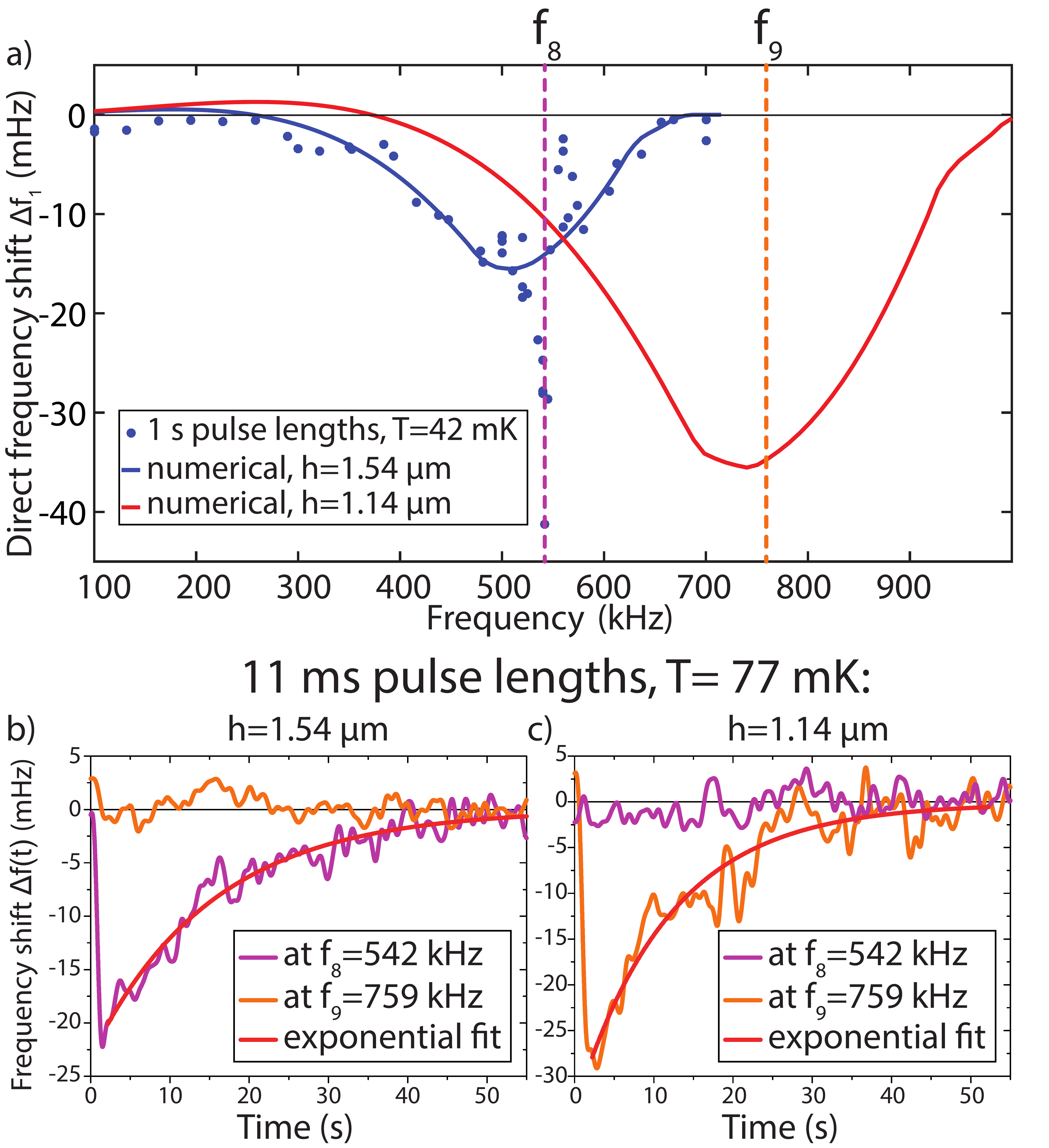}
\caption{\textbf{a)} Frequency shift of the cantilever's first mode at $3$ kHz after applying an rf pulse of $0.2$ mA and $1$ s length using the rf wire. In blue the data \cite{Wagenaar2016} measured at a height of $1.54$ $\micro$m and a temperature of $42$ mK. The solid line is numerically calculated (Eq. \ref{eq:summation}) using a resonant slice thickness of $38$ nm. In red is the simulated signal when the cantilever is positioned $400$ nm closer to the surface. The dotted lines indicate the positions of the higher modes $f_8=542\pm1$ kHz and $f_9=759\pm1$ kHz. \textbf{b)} Frequency shift as function of time after applying an rf pulse at the frequency of the two higher modes at $T=77$ mK. The pulse time of $11$ ms is too short to saturate the spins directly with the oscillating magnetic field of the rf line, but the excitation of the higher modes give a large enough rf field. As expected from a), no signal is observed at the mode of $759$ kHz, while there is for the mode at $542$ kHz. \textbf{c)} The same as in b) but the cantilever has been moved $400$ nm closer to the sample. The mode at $759$ kHz now corresponds to a resonant slice that lies within the sample, resulting in a frequency shift. While the signal at $542$ kHz is greatly reduced. Note that figures b and c are measured at slightly larger temperatures, and have much shorter pulse lengths than a.  }  
\label{fig:fig3}
\end{figure}				
				
\subsection{Results}
Figure \ref{fig:fig3}a shows the frequency shift $\Delta f_1$, according to equation \ref{eq:fitting}, after applying a $1$ second pulse with the rf wire\cite{Wagenaar2016}, at a temperature of $42$ mK. We use a current of $I=0.2$ mA, resulting in an alternating rf field of $B_{rf}=(\micro_0I/2\pi r)\approx5.8$ $\micro$T. The magnetic field strength is enough for a saturation parameter which is at least $s\equiv \gamma^2 (B_{rf}/2)^2 T_1 T_2\approx 30$, which indicate that for sufficient long pulse times the spins will saturate at least $95\%$. However, the rf field generated leads to a flipping rate of approximately $\gamma B_{rf}/4\pi\approx 30$ Hz, which is not fast enough to saturate the spin levels within milliseconds. The solid blue line in the figure is the calculated signal with a height above the sample $h=1.54$ $\micro$m and slice thickness $d=38$ nm. The yellow curve is the simulated signal with the same slice thickness, but at a different height, $400$ nm closer to the surface. The dashed vertical lines are the indicated positions of the $8^{\text{th}}$ and $9^{\text{th}}$ mode of the cantilever. Visible in blue is that we obtain a much larger signal when the radio frequency applied corresponds to the $8^{\text{th}}$ mode, which we attribute to a much larger $B_{rf}$ field, since the resonant slice thickness grows at sufficiently large rf field strengths\cite{Wagenaar2016}.  

To confirm that the larger rf fields are generated by the rotation of the magnetic particle when the higher mode is driven by the rf wire, we measure the signal for much shorter rf pulses. At a pulse length of $11$ ms no frequency shifts are found, except when the frequency is equal to a higher mode and when the frequency of this mode has a resonant slice in the sample. In figures \ref{fig:fig3}b and \ref{fig:fig3}c, typical time traces are shown, after a $11$ ms rf pulse with a frequency that matches exactly the $8^{\text{th}}$ and $9^{\text{th}}$ mode at a temperature of $T=77$ mK. The pulse time of $11$ ms is too short to saturate the spins directly with the $B_{rf}$ of the wire. We see that the $9^{\text{th}}$ mode does not give any signal at $h=1.54$ $\micro$m, but after approaching the cantilever to the sample by $400$ nm, we obtain a clear signal, while the signal at the $8^{\text{th}}$ mode vanishes.  We average multiple time traces, with a low pass filter with a cutoff frequency of $1$ Hz to smooth the data. 
\begin{figure}[t]
\includegraphics[width=\columnwidth]{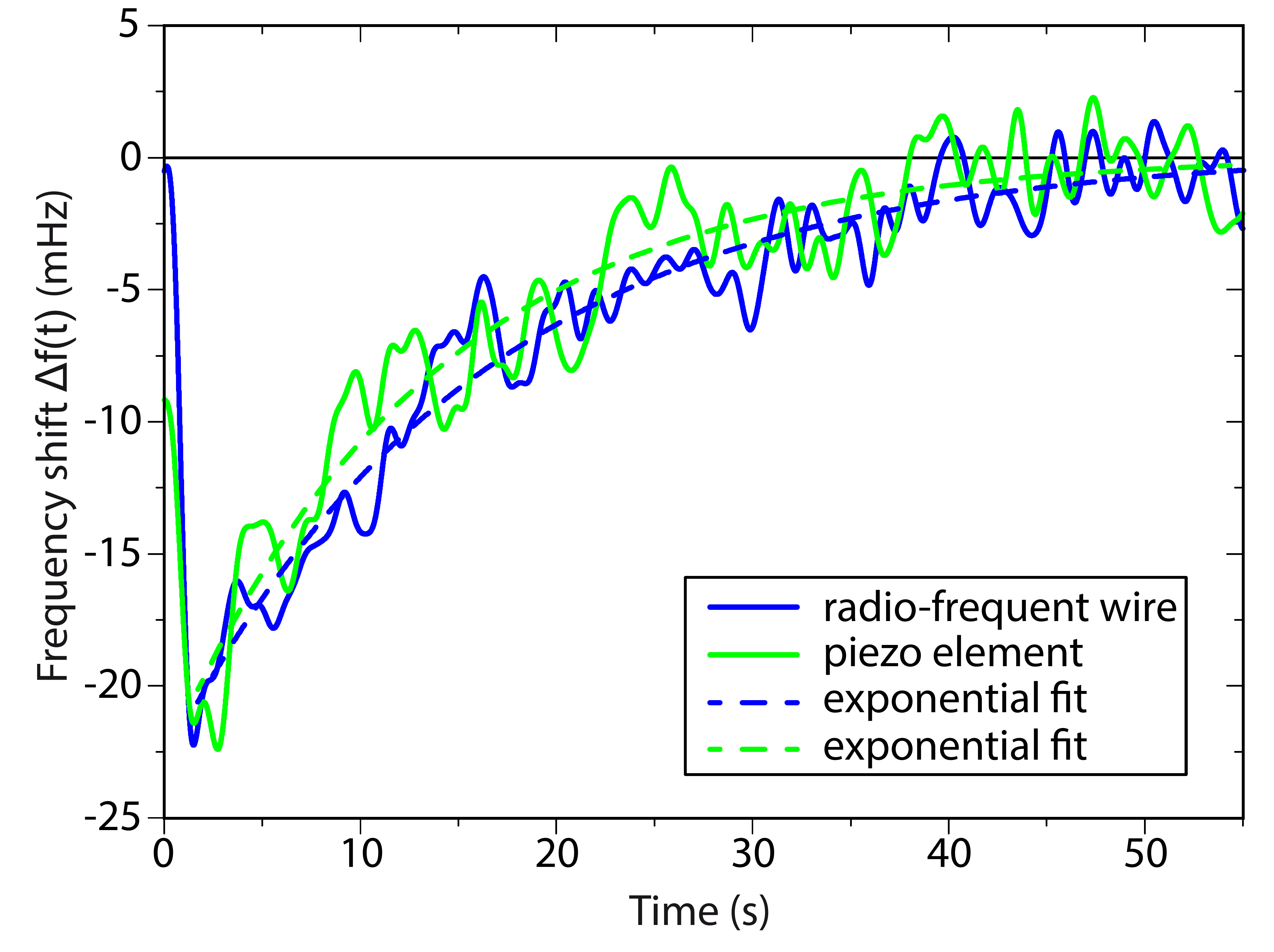}
\caption{Exciting the cantilever with the rf line (blue) with a current of $0.2$ mA and pulse length of $11$ ms, and with the piezo (green) on the cantilever's mount stage with a pulse length of $4$ ms.}
\label{fig:fig4}
\end{figure}
The relaxation times fitted in figures \ref{fig:fig3}b and \ref{fig:fig3}c ($T_1\approx15$ s and $T_1\approx12$ s) with the temperature $T=77$ mK, result in $TT_1\approx 0.9 - 1.2$ sK, in agreement with the well known bulk measurement of  $TT_1=1.2$ sK \cite{Pobell2007}. In the previous report\cite{Wagenaar2016}, this relation was measured for a wider range of temperatures, also for the data collected at the frequency at the $8^{\text{th}}$ mode, showing the same relation.

\begin{figure}[b]
\includegraphics[width=\columnwidth]{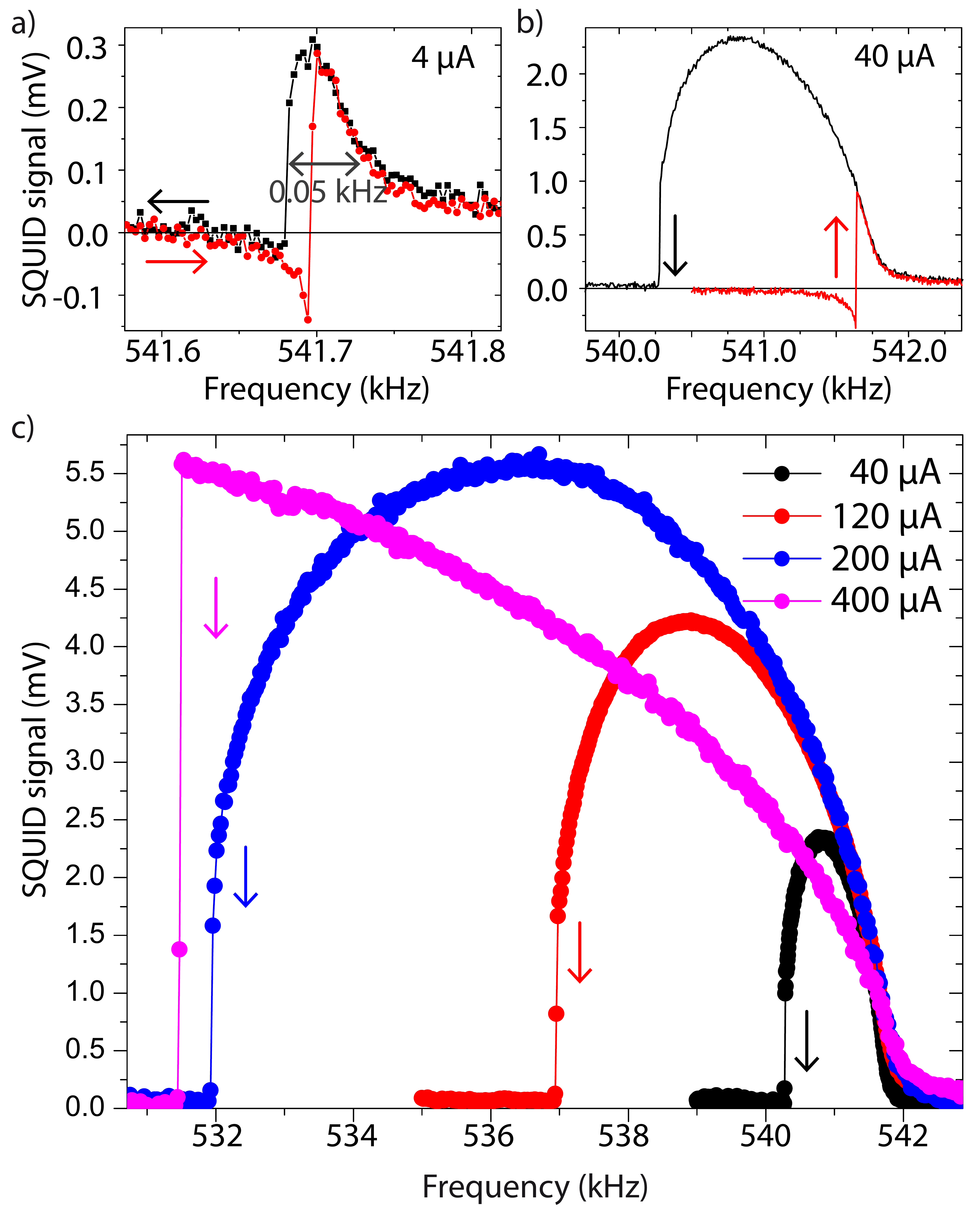}
\caption{\textbf{a) and b)} Exciting the cantilever at its $8^{\text{th}}$ mode, far from the surface, by sweeping the frequency by using the rf wire while recording the SQUID's signal. Already at a current of $40$ $\micro$A, a huge non-linear effect (Duffing) is present. A background signal is subtracted for clarity. \textbf{c)} The sweeps from right to left at different currents. }
\label{fig:fig5}
\end{figure}

Figure \ref{fig:fig4} shows two averaged time traces for two different methods to excite the higher mode. We use the superconducting rf wire to excite the cantilever. The cantilever can be positioned far from the rf wire, enabling experiments on a much larger area than in previous MRFM experiments\cite{Degen2009}. On top of this, no significant heating is present within this experiment, since very low rf currents are already sufficient for driving the cantilever (see Fig. \ref{fig:fig5}). 

Furthermore, we report on experiments in which we do not use the rf wire as a driving mechanism, but solely drive the cantilever using the piezo to excite the fundamental cantilever motion in the phase-locked loop mode. Here, we followed the same measurement procedure, but instead of applying the pulse using the wire, the signal is added to the piezo. We excited the piezo at $f_8=542$ kHz for a very short period of $4$ ms with a voltage of $200$ mV, and did not see an increase in frequency shift when applying larger voltages. Figure \ref{fig:fig4} shows an equal frequency shift was obtained for both methods, which could be an indication of an intrinsic limitation of the rotation angle of the higher modes.
To investigate the possible limitations of the amplitude of the rf fields generated by the higher modes, we need to investigate the maximal  rotation angle $\alpha$ that can be obtained. Figure \ref{fig:fig5} shows the SQUID signal while sweeping the frequency of the current through the rf wire, with the cantilever positioned far from the surface. We see that even for small rf currents the higher modes behave non-linear (Duffing), visible by the hysteresis and typical foldover effect\cite{Duffing1918} when performing a frequency sweep from left to right and vice versa. For $4$ $\micro$A, we have minimal hysteresis, enabling us to estimate the quality factor on the order of $Q_8\approx 10^4$.  

We can give a lower bound for the field strengths by observing the amplification we see of the signal at the frequency of the higher mode and compare it with the signal which you would have expect using only the rf wire. Figure \ref{fig:fig3}a shows that the amplification is almost a factor $3$. The resonant slice thickness is proportional to the rf field strength in the saturation regime\cite{Wagenaar2016}, neglecting possible spin diffusion. From this we obtain a lower bound of approximate $18$ $\micro$T. By using Eq. \ref{eq:RF}, we calculate that this field strength corresponds to  a rotational angle of only $\alpha_0\approx0.01$ degrees. This angle corresponds to an amplitude of the mode's antinodes of only $\delta x_8\approx \frac{\alpha_0 L}{10}\approx1$ nm, with $L$ the length of the cantilever, and $\alpha_0$ in radians.

\section{Discussion and outlook}
Saturation experiments do not need large rf fields, so the rf fields we mechanically generated are enough for measurements of the nuclear spin-lattice relaxation times in condensed matter systems. However, if one would use this technique for high resolution three-dimensional imaging, one needs much larger rf fields. The reason for this is that, in order to exploit the MRFM fully, one would like to flip the spin exactly twice every cantilever period. For example for protons, magnetic field strengths of several mT are needed to use this protocol\cite{Poggio2007}. We see that we obtain at least $18$ $\micro$T, probably hampered by the huge hysteresis of the higher mode. So far, we have only been able to generate at least $18$ $\micro$T at the sample, probably hampered by the huge non linearity of the mode. However, this field was achieved at a large separation between the cantilever and the surface to reduce the impact of the eddy currents which ruin the quality factor of the fundamental mode at $3$ kHz.  When we perform the experiment directly above the sample, we gain a factor $6$.  When using smaller magnetic particles of focused ion beam milled FePt\cite{Overweg2015}, we gain an additional factor of $8$. This can be further improved by using magnets with even larger field gradients\cite{Tao2016}. We believe that these improvements would bring the intensity of the rf field up to the mT range, which would allow the adiabatic rapid passages needed for high resolution three dimension MRFM imaging\cite{Poggio2007}. 

\section{Conclusions}
To summarize, we have proposed a novel mechanical method to generate rf fields using the higher modes of a cantilever. This method can be used within MRFM experiments while using the same cantilever also as force sensor by measuring the frequency shift of the much softer first resonant mode. Our method enables MRFM without being limited by being too close to an rf source. Furthermore, it prevents the usual heating of sample and cryostat at milliKelvin temperatures. We have shown that the field strength can be extended to the mT-regime by using higher field gradient magnets, allowing this technique to be used in the more advanced imaging methods in MRFM. All together, we believe that this new method constitutes a major contribution towards the development of MRFM at ultra-low temperatures.

\section*{Acknowledgements}
We thank F. Schenkel, J. P. Koning, G. Koning and D. J. van der Zalm for technical support. This work was supported by the Dutch Foundation for Fundamental Research on Matter (FOM), by the Netherlands Organization for Scientific Research (NWO) through a VICI fellowship to T.H.O, and through the Nanofront program.

\bibliography{bibliography}

\end{document}